\documentclass[10pt,twocolumn,nofootinbib,prd]{revtex4}
\usepackage{mathrsfs}
\usepackage{graphicx}
\usepackage{amssymb,amsmath,amsthm}
\usepackage{epstopdf}
\usepackage{bm}
 \textwidth = 6.5 in \textheight = 9 in \oddsidemargin =
0.0 in \evensidemargin = 0.0 in \topmargin = 0.0 in \headheight =
0.0 in \headsep = 0.0 in \voffset=0.0 in
\def\ba{\begin{eqnarray}}
\def\ea{\end{eqnarray}}
\def\nn{\nonumber}

\def\D{\mathcal{D}}

\def\:{\boldsymbol{:}}

\def\Tr{\textrm{Tr}}

\long\def\symbolfootnote[#1]#2{\begingroup%
\def\thefootnote{\fnsymbol{footnote}}\footnote[#1]{#2}\endgroup}

\begin{document}

\title{Decoupling of Heavy Kaluza-Klein Modes In Models With Five-Dimensional Scalar Fields}\thanks{This work was supported by the US department of energy.}


\author{Ratindranath Akhoury}
\email{akhoury@umich.edu}
\author{Christopher S. Gauthier}
\email{csg@umich.edu}
\affiliation{Michigan Center for Theoretical Physics \\ Randall Laboratory of Physics \\ University of Michigan, Ann Arbor, MI 48109-1120, USA}

\begin{abstract}
We investigate the decoupling of heavy Kaluza-Klein modes in $\phi^{4}$ theory and scalar QED with space-time topology $\mathbb{R}^{3,1} \times S^{1}$. We calculate the effective action due to integrating out heavy KK modes. We construct generalized RGE's for the couplings with respect to the compactification scale $M$. With the solutions to the RGE's we find the $M$-scale dependence of the effective theory due to higher dimensional quantum effects. We find that the heavy modes decouple in $\phi^{4}$ theory, but do not decouple in scalar QED. This is due to the zero mode of the 5-th component of the 5D gauge field $A_{5}$. Because $A_{5}$ is a scalar under 4D Lorentz transformations, there is no gauge symmetry protecting it from getting mass and $A_{5}^{4}$ interaction terms after loop corrections. In light of these unpleasant features, we explore $S^{1}/\mathbb{Z}_{2}$ compactifications, which eliminate $A_{5}$, allowing for the heavy modes to decouple at low energies. A set of RGEs with respect to the compactification scale are derived for the couplings in scalar QED on the orbifold. It is found that both couplings become weaker as the compactification scale increases. We also explore the possibility of decoupling by including higher dimensional operators. It is found that this is possible, but a high degree of fine tuning is required.
\end{abstract}

\maketitle

\section{Introduction}
Kaluza-Klein models have enjoyed much attention from physicists over the years. Although there is no experimental evidence to suggest the existence of extra hidden dimensions, theorists have continued to entertain the idea because of the promise these models hold. The reason that these extra dimensions have escaped detection so far has traditionally been explained by assuming that these dimensional are curled up so tightly they are too small to probe with present day accelerators. 

A common feature of all Kaluza-Klein models is the existence of an infinite ``tower'' of progressively massive particles. Because these heavy Kaluza-Klein (KK) modes have escaped detection, their effects must decouple at low energy scales \cite{Appelquist:1974tg}. In general Kaluza-Klein models the mass of the lightest heavy KK mode is $M \sim (\textrm{Size of compact dimensions})^{-1}$. If $M$ is higher then the energy scale under consideration, the effective field theory will be a theory of the zero KK mode fields in four dimensions. The heavy KK modes disappear in the low energy limit of the theory. The only effect the heavy modes have on the low energy dynamics are loop-corrections to local operators of the zero KK mode fields, which can be absorbed into the existing couplings and masses of the tree level theory.

In this paper we will investigate the issue of decoupling in Kaluza-Klein models. We focus on the higher dimensional generalizations of $\phi^{4}$ theory and scalar QED. In our analysis, the energy scale under consideration is assumed to be much less than the Planck scale, rendering quantum gravitational effects unimportant. Even at compactification scales far below $M_{pl}$ it is still possible that extra dimensions exist and have discernible effects in our low energy world \cite{Antoniadis:1990ew}.  For much of this paper we will assume that space-time is 4+1 dimensional with an extra spatial dimension that has been periodically identified: $x^{5} \sim x^{5} + 2 \pi R$, where $R$ is the radius of the $5$-th dimension. The resulting space-time manifold has a $\mathbb{R}^{3,1} \times S^{1}$ topology.

For any bosonic field $\bar{\Phi}$ to be well defined on this type of manifold, the field must obey periodic boundary conditions in the compact direction: $\bar{\Phi}(x^{\mu} , x^{5}) = \bar{\Phi}(x^{\mu},x^{5} + 2 \pi R)$. In this report a bar over fields and coordinates denotes the $4+1$ dimension fields and coordinates, while those without the bar denotes their dimensionally reduced counterparts. Indices $M,N,...,$ etc. take values over the total number of compact and non-compact dimensions while $\mu,\nu,..,$ etc. will denote indices over the non-compact dimensions. Since $\bar{\Phi}$ satisfies periodic boundary conditions, it can be expressed as a Fourier series in the compact direction
\begin{gather}
\bar{\Phi}(\bar{x})
=
\sum_{n=-\infty}^{\infty} \Phi^{(n)}(x) e^{i n M \theta}
\end{gather}
where $\theta$ is an angular coordinate patch on the compact dimension and $M = R^{-1}$ is the compactification scale. The action of this field can now be reduced to a 4D action by integrating over the compact dimension. The result is tho action of an infinite number of coupled particles $\Phi^{(n)}$, indexed by the magnitude of their compact momenta. 

In general, after integrating over the fifth dimension, the 5D action becomes the 4D action of a light field sector containing the zero KK modes, a heavy sector containing the nonzero KK modes and an interaction term that connects the two:
\begin{gather}
S^{(\textrm{5d})}
\rightarrow
S_{0}^{(\textrm{4d})}
+
\sum_{n=1}^{\infty}
S_{n}^{(\textrm{4d})}
+
S_{\textrm{int}}^{(\textrm{4d})}.
\end{gather}
At energies below $M$ the heavy modes decouple. The interaction term creates Feynman diagrams with heavy mode loops that make corrections to the masses and couplings of the light sector. These corrections will come in the form of an integral over the  4D dimensional momentum and a sum over the heavy KK modes:
\begin{gather}
\sum_{n=1}^{\infty}
\int \frac{d^{4} p}{(2 \pi)^{4}} f_{n}(p).
\end{gather}
Many procedures have been suggested for regularizing expressions similar to the above \cite{GrootNibbelink:2001bx} \cite{DiClemente:2001ge, Contino:2001uf}, \cite{Delgado:2001ex} , \cite{Contino2:2001gz} \cite{Ghilencea:2001ug,Ghilencea:2001bv,Ghilencea:2001bw}. There is a ongoing debate as to which procedure makes the most sense physically \cite{Alvarez:2006sf}. We do not attempt to answer this question, and we will simply choose the procedure that makes the most physical sense to us. In the context of our discussion, the corrections are treated as a sum of 4D loop corrections originating from an infinite tower of massive particles. Therefore, we choose to evaluate the 4D integral first using dimensional regularization, and then perform the sum over KK modes using zeta function regularization \cite{DiClemente:2001ge}.

The extension of Kaluza-Klein theories to gauge fields is similar to scalar fields, but with some additional subtleties. The first of these is gauge fixing. It is a straightforward task to generalize the gauge fixing done in four dimensions to five dimensions \cite{Dienes:1998vg,Muck:2001yv}. The biggest difference between scalar and gauge fields in Kaluza-Klein models is the additional scalar one obtains after dimensional reduction. The extra component $A_{5}$ of the 5D gauge field becomes a KK tower of 4D Lorentz scalars after dimensional reduction. Being unprotected by any gauge symmetry in 4D, corrections by heavy modes can create new local operators involving the $A_{5}$ zero mode that do not originate from any local operator in the 5D theory.

This paper is organized as follows. In section \ref{LPEFGeneralFormalism} we review the formalism behind the light-particle effective action (LPEA). We use the method first described by \cite{Weisberger:1981xe} to derive our results for the LPEA. In section \ref{ExampleMassivephiTheory} we derive the heavy mode corrections to the mass and coupling in a simple $\phi^{4}$ theory. Here we provide more details on how the LPEA is derived including a discussion about how zeta function regularization is used to control the divergent KK sums. In section \ref{5dScalarQEDTitle} we derive the LPEA of 5D scalar QED. We discuss the effects that the heavy modes have on the low energy observables of the theory and point out that it has a number of unpleasant features. A particularly striking one is a violation of charge universality at the one-loop level. We conclude the section with a discussion of the different ways in which these problems can be eliminated. Finally, in section \ref{Conclusions} we conclude our paper with a summary of our main results.


\section{Light-Particle Effective Action: General Formalism}
\label{LPEFGeneralFormalism}
The following discussion is based on the technique described in \cite{Weisberger:1981xe}. Consider a general action $S[\phi,\chi]$, where $\phi$ represents a ``light'' field whose mass is much smaller compared to the mass of the ``heavy'' field $\chi$. We wish to find the effective action of the light field $\phi$ obtained by integrating out the heavy field $\chi$. The generating functional for the full theory is given by
\begin{gather}
i Z[j,J]
=
\log \int [\D \phi \D \chi] e^{
i S[\phi,\chi] +i \int ( j \phi + J \chi)}
\end{gather} 
here $j$ and $J$ are classical sources for the fields $\phi$ and $\chi$, respectively. In order to get the proper low energy effective field theory, define the {\it light-particle effective action} $\bar{\Gamma}$ as the Legendre transform of $Z$ with respect to only the light particle current.
\begin{gather}
\bar{\Gamma}[\phi_{c},J]
=
Z[j,J]
-
\int j \phi_{c}.
\end{gather}
The functional $\bar{\Gamma}$ generates all diagrams that are 1PI with respect to the light field $\phi$ but not 1PI with respect to the heavy field $\chi$. The light particle effective action, therefore, includes all corrections to the couplings and masses from diagrams containing heavy internal loops. For practical purposes, the steepest descent method can be used to find an approximate expression for $Z$, which in turn can be used to find $\bar{\Gamma}$. The result for the light-particle effective action is:
\begin{gather}
\bar{\Gamma}[\phi_{c}]
=
S[\phi_{c},\chi(\phi_{c})]
+
\frac{i}{2} \Tr\left[
K(\phi_{c}) K^{-1}(0)
\right]
\label{EffectivePot}
\end{gather}
where $\chi(\phi_{c})$ is the solution to the functional equation $\frac{\delta \Gamma}{\delta \chi}=0$ (note that $\Gamma$ is the full 1PI effective action) and $\phi_{c}$ is a c-number background field corresponding to the field operator $\phi$. $K$ is a matrix valued functional of $\phi_{c}$ defined as:
\begin{gather}
K(\phi_{c})
=
\left(
\begin{array}{cc}
\frac{\delta^{2} S}{\delta \phi \delta \phi}[\phi_{c},\chi(\phi_{c})] & \frac{\delta^{2} S}{\delta \phi \delta \chi}[\phi_{c},\chi(\phi_{c})] \\
\frac{\delta^{2} S}{\delta \chi \delta \phi}[\phi_{c},\chi(\phi_{c})] & \frac{\delta^{2} S}{\delta \chi \delta \chi}[\phi_{c},\chi(\phi_{c})]
\end{array}
\right).
\end{gather}
The generalization of this procedure to an infinite number heavy modes is clear. Suppose we have a countably infinite number of sectors in a theory, each describing the dynamics of the fields $\Phi^{(n)}$, where $\Phi^{(n)}$ is a shortened notation for a finite set of fields that each have the same KK mass term $n^{2} M^{2}$. The fields $\Phi = \Phi^{(0)}$ have no KK mass and are therefore considered light. Since in most cases $\Phi^{(n)}(\Phi_{c}) = 0$, off-diagonal terms in the $K$-matrix between different KK mode sectors vanish:
\begin{gather}
{\scriptsize
K(\Phi_{c})
=
\left(
\begin{array}{ccccc}
\ddots & & & & \\
            &K_{1}(\Phi_{c}) & & & \\
            & & K_{0}(\Phi_{c}) & & \\
            & & & K_{-1}(\Phi_{c}) & \\
            & & & & \ddots
\end{array}
\right)}
.
\end{gather}
In this case the $K$-matrix has a block diagonal form. The diagonal entries are square matrices defined as $K_{n} = \frac{\delta S}{\delta \Phi^{(n)} \delta \Phi^{(-n)}}$. Since $K_{n} = K_{-n}$ the light-particle effective action can now be written as
\begin{gather}
\bar{\Gamma}[\Phi_{c}]
=
S[\Phi_{c}] + i\sum_{n=1}^{\infty} \Tr\left[
 K_{n}(\Phi_{c}) K_{n}(0)^{-1} \right].
\end{gather}
Note that we have ignored light loop diagram contributions by excluding the $n=0$ term in the sum. In many discussions of Kaluza-Klein models \cite{Cheng:2002iz}, \cite{Puchwein:2003jq} the zero modes are included in loop corrections. We take a different approach, and investigate the low energy effective field theory of the light zero mode fields due to loop corrections from the heavy KK mode fields. We will see that this approach leads to new divergences and a non-trivial dependence of the masses and couplings on the compactification scale $M$.

\section{Massive $\phi^{4}$ Theory}
\label{ExampleMassivephiTheory}
To see how decoupling of massive KK modes can effect low energy physics let us first consider the simple model: $\phi^{4}$ theory. The action in 5D is
\begin{gather}
\int d^{5}\bar{x} 
\left[
\frac{1}{2} \partial_{M} \bar{\phi}  \partial^{M} \bar{\phi}
-
\frac{1}{2} \bar{m}^{2} \bar{\phi}^{2}
-
\frac{\bar{\lambda}}{4!}
\bar{\phi}^{4}
\right].
\end{gather}
We will assume in our analysis that ``low energy'' means around the electroweak scale: $\sim 1$ TeV, well below the Planck scale, where gravity can be considered classical and the quantum field theory description is still valid. After dimensional reduction the result is
\begin{widetext}
\small \begin{gather}
\int d^{4}x
\left(
 \frac{1}{2} \sum_{n=-\infty}^{\infty}[\partial_{\mu}\phi^{(n)}  \partial^{\mu}\phi^{(-n)}
-
(m^{2} + n^{2} M^{2})
\phi^{(n)} \phi^{(-n)}]
-
\frac{\lambda}{4!}
\sum_{k,l,m=-\infty}^{\infty} \phi^{(k)} \phi^{(l)} \phi^{(-m)} \phi^{(m-k-l)}
\right)
\label{dimRedUnbroken}
\end{gather}
\end{widetext}
where $m^{2}= \bar{m}^{2}$ and $\lambda = \frac{\bar{\lambda}}{2 \pi R}$. Notice that this describes the action of a infinite number of increasingly massive scalars, coupled through a quartic interaction term. The interaction term allows for the heavy modes to appear in loop corrections to the zero mode mass $m$ and 4D coupling $\lambda$. If we consider the case where $m \ll M$ the theory can be separated into a ``light'' part containing the field $\phi = \phi^{(0)}$ and a heavy sector containing all $\phi^{(n)}$ with $|n|>0$. To one-loop order, interactions between different nonzero modes will be negligible. Therefore, the important part of the interaction in (\ref{dimRedUnbroken}) will be

\begin{gather}
\frac{\lambda}{4!}\sum_{k,l,m=-\infty}^{\infty}\phi^{(k)} \phi^{(l)} \phi^{(-m)} \phi^{(m-k-l)}
\rightarrow
\nn \\
\frac{\lambda}{4!} \phi^{4} + \frac{\lambda}{2} \sum_{k=1}^{\infty} \phi^{2} \phi^{(k)} \phi^{(-k)}.
\end{gather}
Using the formula in (\ref{dimRedUnbroken}) and dimensionally regularizing the 4D loop integrals, we find that the corrections to the effective potential are   
\begin{widetext}
\begin{gather}
\delta V_{eff}(\phi) 
= 
\frac{\lambda}{2}
\left(
\sum_{n=1}^{\infty}
\frac{\Gamma(1 - \frac{d}{2})}{(4 \pi)^{d/2}}
\frac{\mu^{4 - d}}{(m^{2} + n^{2} M^{2})^{1-d/2}}
\right)
\phi^{2}
-
\frac{3 \lambda^{2}}{4!}
\left(
\sum_{n=1}^{\infty}
\frac{\Gamma(2 - \frac{d}{2})}{(4 \pi)^{d/2}}
\frac{\mu^{4 - d}}{(m^{2} + n^{2} M^{2})^{2-d/2}}
\right)
\phi^{4}.
\end{gather}
\end{widetext}
The integrals over 4D momenta have been dimensionally regularized and $\mu$ is an arbitrary mass parameter needed to keep the result dimensionally correct. The sum in this expression is clearly divergent and also requires regularization. 

When regularization of a divergent sum is needed the most natural technique for doing so is zeta function regularization \cite{Hawking:1976ja}. In our case, we start by defining a generalized zeta function $\zeta_{L}(s)$ for $s<1$ as
\begin{gather}
\zeta_{L}(s) = \sum_{n=1}^{\infty} [m^{2} + n^{2} M^{2}]^{-s}.
\label{GenZeta}
\end{gather}
It is easy to show that the quantum corrections to the potential in terms of this generalized zeta function are
\begin{gather}
\delta V_{eff}(\phi)
=
\lambda (4 \pi \mu^{2})^{\epsilon/2}
\frac{\Gamma(\frac{\epsilon}{2} - 1) \zeta_{L}(\frac{\epsilon}{2} - 1)}{32 \pi^{2}}
\phi^{2}
\nn \\
-
\frac{3 \mu^{\epsilon} \lambda^{2}}{4!}
(4 \pi \mu^{2})^{\epsilon/2}
\frac{\Gamma(\frac{\epsilon}{2} ) \zeta_{L}(\frac{\epsilon}{2})}{16 \pi^{2}}
\phi^{4}.
\end{gather} 
The problem now becomes to analytically continue $\zeta_{L}(s)$ to an entire function on the complex plane. The strategy is to now rewrite our summation expression for $\zeta_{L}$ in terms of the Riemann zeta function, which has a well known analytic continuation on $\mathbb{C}$. The generalized zeta function (\ref{GenZeta}) written in terms of the Riemann zeta function $\zeta$ is 
\begin{gather}
\zeta_{L}(s)
=
M^{-2 s} \sum_{k = 0}^{\infty} (-1)^{k} \alpha(s,k) X^{2 k} \zeta(2 s + 2 k)
\end{gather}
where $X^{2} = \frac{m^{2}}{M^{2}}$ and
\begin{gather}
\alpha(s,k)= \lim_{x\rightarrow s}\frac{\Gamma(x+k)}{\Gamma(k+1) \Gamma(x) }
\qquad
\textrm{for $k \in \mathbb{Z}$.}
\end{gather}
One might worry that the sum over $k$ might diverge. Fortunately, at $s=-1$ and $s=-2$ the sum truncates, since $\alpha(s,k)$ vanishes for all $k \geq - s + 1$ when $s = - n$ where $n \in \mathbb{N}$. The final result for the bare coupling $\lambda_{b}$ and mass $m_{b}$ to one loop order are\footnote{In the report we have used $\frac{1}{\epsilon} \left(\frac{\mu}{M}\right)^{\epsilon}$ as a short hand for $\frac{1}{\epsilon} - \log \left[\frac{M}{\mu}\right]$. The correspondence is not exact, but it is acceptable since we are only concerned with the divergent and log parts of the corrections.}
\begin{gather}
\lambda_{b} 
= 
\mu^{\epsilon}
\lambda
\left( 
1
-
\frac{3 \lambda}{16 \pi^{2} \epsilon}
\left(
\frac{\mu}{M}
\right)^{\epsilon}
\right)
\\
m_{b}^{2}
=
m^{2} 
-
\frac{\lambda m^{2}}{32 \pi^{2} \epsilon}
\left(
\frac{\mu}{M}
\right)^{\epsilon}
- 
\frac{\lambda M^{2} \zeta(3)}{64 \pi^{4}}.
\label{lambdaANDmassCorrection}
\end{gather}
Here we have denoted the $\lambda$ and $m$ as the renormalized coupling and mass. Note that the divergent $\frac{1}{\epsilon}$ term does not depend on the compactification scale, so one can define the renormalized coupling or mass for all values of the scale $M$, simultaneously. It should also be noted that the heavy modes have decoupled since all of their effects can be absorbed into the original masses and couplings from the tree level theory. The renormalized parameters depend on two scales: the 4D mass scale $\mu$, and the compactification scale $M$. The presence of an $M^{2}$ term in the $m^{2}$ correction is similar to the quadratic divergence present in a crude cutoff regularization of the scalar mass squared. This $M^{2}$ piece is an unavoidable consequence of adding a new scale to a theory with no 4D symmetries to protect it.

We are interested in how the theory depends on the compactification scale $M$ since it determines how the effective theory will change as the size of the compact dimension changes. In some sense $M$ acts like a renormalization scale for the 5D  momenta, since its inverse $M^{-1} = R$ determines the 5D length scale. Keep in mind, that as they are defined, the bare parameters are independent of $M$. If we treat the compactification scale in the same way one treats an arbitrary mass scale, we can construct a RG equation for $\lambda$ with respect to $M$:
\begin{gather}
\frac{d \lambda}{d \log M}
=
- \frac{3 \lambda^{2}}{16 \pi^{2}}.
\label{ScalarRGE}
\end{gather}
This has the familiar form of an RG equation, but the scale that it is with respect to has a very different interpretation; the scale $M$ is not an arbitrary adjustable mass scale but an intrinsic scale of the space-time topology. The solution to this equation is
\begin{gather}
\lambda(M)
=
\frac{\lambda_{0}}{1 + \frac{3 \lambda_{0}}{16 \pi^{2}} \log M/M_{0}}.
\label{lambdaRGESol}
\end{gather} 
The value $\lambda_{0}$ of the coupling at the scale $M_{0}$ is arbitrary and must be set by experiment or other theoretical considerations.

The sign of the beta function in (\ref{ScalarRGE}) indicates that the coupling is asymptotically free with respect to the compactification scale. The heavy modes have a screening effect that becomes more pronounced as the scale $M$ increases. As the compactification scale decreases, the screening effect becomes weaker and the coupling becomes stronger. To one loop order, the coupling appears to have a Landau pole in the IR region. We can use this Landau pole to trade the dimensionless coupling $\lambda$ in for a dimensionful scale $\Lambda$ where the perturbative expansion breaks down: 
\begin{gather}
\lambda(M)
=
\frac{16 \pi^{2}}{3}
\frac{1}{\log( M / \Lambda)}.
\end{gather}
In principle, the scale $\Lambda$ represents the lowest scale that we can use before perturbation theory fails. In order for this analysis to work, $\Lambda$ must be assumed to be much lower then the true compactification scale $M$. Whether or not this Landau pole actually exists or is an artifact of perturbation theory is not clear. This issue of triviality in the IR limit remains an open question. 

This simple example illustrates the effective 4D description of a simple 5D scalar field model. The issues that plague this model are no more serious then those that effect 4D $\phi^{4}$ theory. However, as we'll see in the next section, adding a 5D gauge field presents new problems when trying to find a sensible 4D description.

\section{5D Scalar QED}
\label{5dScalarQEDTitle}
Now consider a more interesting model, scalar QED in 5D with the fifth dimension compactified on $S^{1}$:
{\small\begin{gather}
\int d^{5}\bar{x} \left[|\D_{M} \bar{\phi}|^{2} 
-
m^{2}
|\bar{\phi}|^{2}
-
\frac{\bar{\lambda}}{3!}
|\bar{\phi}|^{4}
-\frac{1}{4} \bar{F}_{MN} \bar{F}^{MN} 
\right].
\label{5dScalarQED}
\end{gather}}
In this model $\bar{\phi}$ is complex and has a charge $\bar{e}$ that couples to a $U(1)$ gauge field $\bar{A}_{M}$. The addition of a gauge field presents new complications to be dealt with, in particular the issue of gauge fixing. We have chosen the 5D Lorentz gauge in order to simplify calculations: 
\begin{gather}
\mathscr{L}_{GF}
=
-\frac{1}{2} (\partial^{\mu} \bar{A}_{\mu} + \partial^{5} \bar{A}_{5})^{2}.
\end{gather}
Ghosts decouple in this model so we need not worry about them. Including the gauge fixing term, the dimensional reduced form of the action (\ref{5dScalarQED}) is
\begin{widetext}
{\footnotesize \begin{gather}
\int d^{4}x
\left[
\sum_{-\infty}^{\infty}
\bigg(
\frac{1}{2}
A_{\mu}^{(n)}
[
\square_{4}
+
n^{2} M^{2}
]A^{\mu (-n)}
-
\frac{1}{2} A_{5}^{(n)}
\left[
\square_{4}
+
n^{2} M^{2}
\right]
A_{5}^{(-n)}
+
\partial_{\mu} \phi^{(n)} \partial^{\mu}\phi^{\ast (-n)}
-
\left[ m^{2} + n^{2} M^{2}\right] \phi^{(n)} \phi^{\ast (-n)} 
\bigg)
\right.
\nn \\
+
\sum_{-\infty}^{\infty}
e
\bigg(
i A_{\mu}^{(n)} 
(\phi^{(m)} \partial^{\mu}\phi^{\ast (-n-m)}
-
\phi^{\ast (m)} \partial^{\mu}\varphi^{(-n-m)})
+
(n + m) M
 A_{5}^{(n)} 
(\phi^{\ast (m)} \phi^{(-n-m)}
-
\phi^{(m)} \phi^{\ast (-n-m)})
\bigg)
\nn \\
+
\sum_{-\infty}^{\infty}
\Bigg(
e^{2}
\phi^{(n)} \phi^{\ast (m)} 
\left(
A_{\mu}^{(k)} A^{\mu (-n-m-k)}
-
A_{5}^{(k)} A_{5}^{(-n-m-k)}
\right)
-
\frac{\lambda}{3!}
\phi^{(n)} \phi^{(m)} \phi^{\ast (k)} \phi^{\ast (-n-m-k)} 
\bigg)
\Bigg].
\end{gather}}
\end{widetext}
We should note that the KK modes of $\bar{A}_{5}$ only get a KK mass from the gauge fixing term, and so the nonzero $A_{5}$ KK modes decouple at low energies. In a more general gauge fixing, the nonzero KK modes will not have a mass, and therefore won't decouple in the low-energy limit. However, in 4D the gauge transformation acts on the KK modes of $A_{5}$ like
\begin{gather}
\delta A_{5}^{(n)}
=
-\frac{ i n}{e} \Lambda^{(n)}
\label{A5Transform}
\end{gather}
where $\Lambda^{(n)}$ is the $n$-th term in the Fourier expansion of the 5D gauge parameter $\bar{\Lambda}$. Therefore, with the exception of the zero mode (which has $\delta A_{5}^{(0)} = 0$), the $A_{5}$ KK modes are unphysical gauge degrees of freedom. So not only does the Lorentz gauge simplify the 4D gauge field propagator, it also eliminates the nonzero $A_{5}$ KK modes from the low-energy theory. While the $n\neq 0$ $A_{5}$ modes are of no consequence, $A_{5}^{(0)}$ is a physical degree of freedom and does not decouple in the low energy theory. We now have to contend with corrections to the mass and couplings of $A_{5}^{(0)}$, which in general will not be the same as those of the $A_{\mu}^{(0)}$. This is not surprising since the former are not protected by the gauge symmetries of the 4D theory.

After a tedious calculation of integrating out the heavy modes using the program described in section \ref{LPEFGeneralFormalism}, and demonstrated in section \ref{ExampleMassivephiTheory}, we are left with the following one-loop corrected action for the zero mode sector.  
\begin{widetext}
\begin{gather}
\int d^{4}x
\left(
-
\frac{\Gamma_{A}}{4}
F_{\mu\nu}
F^{\mu\nu}
-
\frac{1}{2} A_{5}
\left[
\Gamma_{5}
\square_{4}
+
\Gamma_{m_{5}^{2}}
\right]
A_{5}
+
\Gamma_{\phi} \partial_{\mu} \phi \partial^{\mu}\phi^{\ast}
-
\Gamma_{m} m^{2} |\phi|^{2}
\right.
\nn \\
\left.
+
i
\Gamma_{e}
e
A_{\mu}
(\phi \partial^{\mu}\phi^{\ast}
-
\phi^{\ast} \partial^{\mu}\phi)
+
\Gamma_{e^{2}}^{(A_{\mu})}
e^{2}
|\phi|^{2}
A_{\mu} A^{\mu}
-
\Gamma_{e^{2}}^{(A_{5})}
e^{2}
|\phi|^{2}
A_{5}^{2}
-
\Gamma_{A_{5}^{4}}
A_{5}^{4}
-
\frac{\lambda \Gamma_{\lambda}}{3!}
|\phi|^{4}
\right)
\end{gather}
where the infinite contributions are:
\begin{gather}
\Gamma_{A} - 1
=
-
\frac{e^{2}}{24 \pi^{2} \epsilon}
\left(
\frac{\mu}{M}
\right)^{\epsilon}
,
\qquad
\Gamma_{5} - 1
=0
,
\qquad
\Gamma_{m_{5}^{2}}
=
\frac{e^{2} m^{2}}{4 \pi^{2} \epsilon}
\left(
\frac{\mu}{M}
\right)^{\epsilon}
+
\frac{3 e^{2} M^{2} \zeta(3)}{8 \pi^{4}}
,
\nn \\
\Gamma_{\phi} - 1
=
\frac{e^{2}}{4 \pi^{2} \epsilon}
\left(
\frac{\mu}{M}
\right)^{\epsilon}
,
\qquad
\Gamma_{m}
=
1
-
\frac{3 e^{2} - 2 \lambda}{24 \pi^{2} \epsilon}
\left(
\frac{\mu}{M}
\right)^{\epsilon}
+
\frac{(\lambda + 6 e^{2}) M^{2} \zeta(3)}{24 \pi^{4} m^{2}},
\nn \\
\Gamma_{e}
=
1
+
\frac{e^{2}}{4 \pi^{2} \epsilon}
\left(
\frac{\mu}{M}
\right)^{\epsilon}
,
\qquad
\Gamma_{e^{2}}^{(A_{\mu})}
=
1
+
\frac{e^{2}}{4 \pi^{2} \epsilon}
\left(
\frac{\mu}{M}
\right)^{\epsilon}
,
\qquad
\Gamma_{e^{2}}^{(A_{5})}
=
1
+
\frac{\lambda + 6 e^{2}}{12 \pi^{2} \epsilon}
\left(
\frac{\mu}{M}
\right)^{\epsilon},
\nn \\
\Gamma_{A_{5}^{4}}
=
\frac{e^{4}}{16 \pi^{2} \epsilon}
\left(
\frac{\mu}{M}
\right)^{\epsilon}
,
\qquad
\lambda\Gamma_{\lambda}
=
\lambda
+
\frac{5 \lambda^{2} - 6  \lambda e^{2} + 72 e^{4}}{24 \pi^{2} \epsilon}
\left(
\frac{\mu}{M}
\right)^{\epsilon}.
\label{ZCorrections}
\end{gather}
\end{widetext}
The new coupling constants $\lambda$ and $e^{2}$ are related to their 5D counterparts by $\lambda = \frac{\bar{\lambda}}{2 \pi R}$ and $e^{2} = \frac{\bar{e}^{2}}{2 \pi R}$. The one-loop corrected action above is invariant under 4D gauge transformations since $\Gamma_{e^{2}}^{(A_{\mu})} = \Gamma_{e}^{2}/\Gamma_{\phi}$. Unlike the effective action found in section \ref{ExampleMassivephiTheory}, the heavy modes have not decoupled in the low energy theory. The heavy modes have made nonzero contributions to the $A_{5}^{2}$ and $A_{5}^{4}$ local operators, operators that were absent in the tree level action. The appearance of these news operators in the low energy effective action should come as no surprise. Looking at (\ref{A5Transform}) it is clear that there is no gauge symmetry in four dimensions that acts on $A_{5}$. Since $A_{5}$ is a scalar with respect to the action of the 4D Poincar\'e group, there is no reason to expect that it will not develop nonzero mass and quartic coupling corrections. It should be noted that the $M^{2}$ piece in the $A_{5}$ mass correction has been calculated previously, though in a different context (\cite{Cheng:2002iz}, \cite{Puchwein:2003jq}). It is now clear that counter terms which respect the original 5D gauge symmetry are not sufficient to render the 4D effective theory finite. Indeed, if one wanted to keep the mass and quartic coupling corrections of the $A_{5}$ finite then the original 5D theory would have to contain a mass and quartic coupling counter term, manifestly breaking the 5D gauge symmetry.

What is also troubling is that the corrections to the $|\phi|^{2} A_{\mu} A^{\mu}$ and $|\phi|^{2} A_{5}^{2}$ vertices are different, which implies that the charge receives a different correction for different vertices. If we were to find the correction to the electric charge by evaluating corrections to the $|\phi|^{2} A_{5}^{2}$ vertex we would find that the one-loop correction to the charge is
\begin{gather}
\delta e^{2}
=
\frac{e^{2}(\lambda + 3 e^{2})}{12 \pi^{2} \epsilon}
\left(
\frac{\mu}{M}
\right)^{\epsilon}
.
\label{chargeA5Correction}
\end{gather}
This is in contrast to the charge correction that is obtained from the $|\phi|^{2} A_{\mu} A^{\mu}$ vertex:
\begin{gather}
\delta e^{2}
=
\frac{e^{4}}{24 \pi^{2} \epsilon}
\left(
\frac{\mu}{M}
\right)^{\epsilon}.
\end{gather}
This has consequences for charge universality since the correction to the charge in (\ref{chargeA5Correction}) has a dependence on $\lambda$ which in turn depends on the matter field $\phi$. Again this is due to the absence of a gauge symmetry acting on $A_{5}$. There are no Ward identities that require $\Gamma_{e^{2}}^{(A_{5})}$ to be equal to $\Gamma_{\phi}$. Thus we have no reason to expect that the correction to $e^{2}$ at the $|\phi|^{2} A_{5}^{2}$ vertex will be the same as that at the $|\phi|^{2} A_{\mu} A^{\mu}$ vertex. In contrast, the local operators involving $A_{\mu}$ do satisfy 4D Ward identities to one-loop order, leading to the equalities: $\Gamma_{\phi} = \Gamma_{e}$ and $\Gamma_{e^{2}}^{(A_{\mu})} = \Gamma_{e}$, which are indeed satisfied by the corrections found in (\ref{ZCorrections}).

It would seem that this theory is inherently ``sick'', leading to several unpleasant features that one would like to be absent from a realistic model. There are two possible ways that this problem can be alleviated, which we will discuss below.

\subsection{$S^{1}$ versus $S^{1}/\mathbb{Z}_{2}$ Compactifications}
The first, and most straight forward way to deal with the breaking of charge universality is to choose compactifications of the extra dimensions that do not permit an $A_{5}$ zero mode. In a $S^{1}/\mathbb{Z}_{2}$ orbifold compactifications \cite{Muck:2001yv} the boundary conditions on the components of the 5D gauge field change to
\begin{gather}
\bar{A}_{M}(x,y) 
=
\bar{A}_{M}(x,y+ 2 \pi n R),
\\
\bar{A}_{\mu}(x,y)
=
\bar{A}_{\mu}(x, - y),
\\
\bar{A}_{5}(x,y)
=
-
\bar{A}_{5}(x, - y).
\end{gather}
These boundary conditions lead to a Fourier series expansion of $\bar{A}_{\mu}$ and $\bar{A}_{5}$ 
\begin{gather}
\bar{A}_{\mu}(x,y)
=
\frac{A_{\mu}^{(0)}(x)}{\sqrt{2 \pi R}}
+
\sum_{n=1}^{\infty}
\frac{A^{(n)}_{\mu}(x)}{\sqrt{\pi R}}
\cos\left(n M y\right)
\\
\bar{A}_{5}(x,y)
=
\sum_{n=1}^{\infty}
\frac{A^{(n)}_{5}(x)}{\sqrt{\pi R}}
\sin\left(n M y\right).
\end{gather}
The twisted boundary condition on $\bar{A}_{5}$ precludes the existence of an $A_{5}$ zero mode. Therefore, there is no additional scalar in the 4D effective theory that will lead to different corrections for the gauge coupling $e$. All the coupling and mass corrections for the remaining zero mode fields are the same as in the $S^{1}$ compactification case, except for a factor of $\frac{1}{2}$ due to the different sums over the KK modes. This is a elegant resolution to the problem with the $A_{5}$ zero mode: get rid of it. The fact that an orbifold compactification eliminates the $A_{5}$ zero mode is a pleasant surprise. Orbifold compactifications are already an attractive possibility since they allow for chiral fields \cite{Dienes:1998vg} and lead to realistic string models \cite{Dixon:1985jw,Dixon:1986jc}


Since the $A_{5}$ zero mode was our only obstacle to obtaining a consistent correction for the charge, we can now write down a set of a RGE's for the 4D couplings $\lambda$ and $e^{2}$ with respect to the scale $M$:
\begin{gather}
\frac{d \lambda}{d \log M}
=
-
\frac{5 \lambda^{2} - 18 \lambda e^{2} + 72 e^{4}}{48 \pi^{2}},
\label{SQEDLambdaRGE}
\\
\frac{d e^{2}}{d \log M}
=
-
\frac{e^{4}}{48 \pi^{2}}.
\end{gather}
It is interesting to note that the coefficient of $e^{4}$ in (\ref{SQEDLambdaRGE}), is in general equal to $18 (d-1)$ where $d$ is the {\it total} number of space-time dimensions. Even though this is an effective theory in four dimensions the couplings still ``feel'' the effects of the 5-th dimension. Qualitatively, the dimensional dependence of the $\lambda$ beta function will not lead to any new effects at different dimensions since the beta function is negative definite for all $d > 1$. Because each of the beta functions are negative definite, we expect these two couplings to be asymptotically free in $M$ just as we found with the scalar coupling in section \ref{ExampleMassivephiTheory}. 

\subsection{Higher Dimensional Operators}
It is well known that the $\phi^{4}$ theory in five dimensions is nonrenormalizable. Thus the 5D high energy theory will in general contain nonrenormalizable operators. In the low energy theory, these new operators can introduce new corrections to the couplings and masses \cite{Oliver:2003cy},\cite{Oliver:2003sb} in such a way as to retain charge universality and the masslessness of $A_{5}$. After some careful study one can show that in order accomplish that, we at least need to include the following operators:
\begin{widetext}
\begin{gather}
\frac{c_{1}}{\Lambda^{2}}
(\D_{M} \D_{N} \bar{\phi})
(\D^{M} \D^{N} \bar{\phi})^{\ast}
,
\qquad
\frac{c_{2}}{\Lambda^{2}}
(\D^{M} \D_{M} \bar{\phi})
(\D^{N} \D_{N} \bar{\phi})^{\ast}
,
\qquad
\frac{- i c_{3}}{\Lambda^{5/2}}
\bar{F}^{MN}
(\D_{M} \bar{\phi})
(\D_{N} \bar{\phi})^{\ast},
\nn \\
\frac{c_{4}}{\Lambda^{3}}
|\bar{\phi}|^{2}
(\D_{M} \bar{\phi})
(\D^{M} \bar{\phi})^{\ast},
\qquad
\frac{c_{5}}{\Lambda^{3}}
|\bar{\phi}|^{2}
\bar{F}_{MN} \bar{F}^{MN}
,
\qquad
\frac{c_{6}}{\Lambda^{4}}
\bar{F}^{MN}
(\D_{L} \D^{M} \D_{M}  \bar{\phi})
(\D^{L} \D^{N} \D_{N} \bar{\phi})^{\ast},
\nn \\
\frac{c_{7}}{\Lambda^{4}}
\bar{F}^{MN}
(\D_{L} \D_{M} \D_{N}  \bar{\phi})
(\D^{L} \D^{M} \D^{N} \bar{\phi})^{\ast},
\qquad
\frac{c_{8}}{\Lambda^{5}}
[(\D_{M} \bar{\phi}) (\D^{M} \bar{\phi})^{\ast}]^{2}
,
\qquad
\frac{c_{9}}{\Lambda^{5}}
\left|(\D_{M} \bar{\phi} \D^{M} \bar{\phi})\right|^{2}
\label{NonRenOp}.
\end{gather}
\end{widetext}
The nine coefficients of these operators can each be adjusted so the divergent corrections to the masses and couplings of the $A_{\mu}$ and $A_{5}$ fields are equal. However, there are several downsides to this approach. For one, the $c_{i}$'s ($i=1,\dots , 9$) in (\ref{NonRenOp}) require a high degree of fine tuning in order to achieve the desired results. The second drawback is that the scale $\Lambda$ where these operators are important, has to be close to the mass $m$ of the zero mode $\phi$. Even higher dimensional nonrenormalizable operators can be neglected if their couplings are set to zero, but again there is the issue of fine tuning.

\section{Conclusion}
\label{Conclusions}
In this paper we have discussed the effects that heavy KK modes can have on the low energy physics. The heavy modes in $\phi^{4}$ theory do decouple in the low energy theory, their only effects being corrections to the existing local operators. The one-loop corrections to the coupling lead to a RGE with respect to the compactification scale $M$. The coupling is found to be asymptotically free with respect to the compactification scale; the coupling strength becoming weak at high scales, or equivalently at very small compactification size.

The extra-dimensional extension of scalar QED lead to a low energy theory where the heavy KK modes did not decouple. The reason behind the non-decoupling is the additional scalar zero mode $A_{5}$ that originated as the 5-th component of the 5D gauge field. Since $A_{5}$ was a 4D scalar it was not protected from receiving mass and quartic terms after heavy loop corrections; terms that were not present in the tree level zero mode sector. Excluding $A_{5}$, the low energy theory is 4D scalar QED, and if not for $A_{5}$ the heavy modes would decouple. Therefore, the most direct route to ensuring the decoupling of the heavy modes is to choose compactifications where the $A_{5}$ zero mode is nonexistent. This is the case for $S^{1}/\mathbb{Z}_{2}$ orbifold compactifications. We found that when the theory is placed on $\mathbb{R}^{3,1} \times S^{1}/\mathbb{Z}_{2}$, the 4D low energy theory is scalar QED, with no trace of the heavy modes except for corrections to the local operators of the tree level action. With the heavy modes decoupled, we can define a unique correction to the electric charge and find the RGE's with respect to the compactification scale $M$ for the scalar coupling $\lambda$ and charge $e$. The $M$-scaling behavior of the two couplings are qualitatively the same as the scalar coupling in 5D $\phi^{4}$ theory, becoming weak when the size of the compact dimension becomes very small. We also discussed the possible role that higher dimensional local operators might have in achieving decoupling. It was determined that although it was possible, the high degree of fine tuning required made it an unattractive resolution to the problem.

Since gauge symmetry plays an important part in the decoupling of heavy modes, it would be worth investigating whether models with additional symmetries can protect $A_{5}$ (or fields like it) from receiving unwanted corrections. A possible extension to our model that would incorporate additional symmetries are non-abelian models. In the non-abelian extension, the $A_{5}$ zero mode becomes a gauge multiplet of 4D scalars in the adjoint representation of the gauge group. The invariance of the $A_{5}$ multiplet under the gauge symmetry might have effects on the one-loop corrections to the $A_{5}$ $n$-point functions. It may also be worthwhile to study heavy mode decoupling in standard fermionic QED. Fermions have additional space-time symmetries that may result in the cancellation of at least some of the corrections to the $A_{5}$ two and four point functions. Randall-Sundrum models may also contain interesting effects from the standpoint of KK mode decoupling. The inclusion of branes might make it possible to gently cancel the unpleasant heavy mode corrections, even if a field similar to $A_{5}$ remains in the low energy theory. 

The generalization of the 5D QED model to include spontaneous symmetry breaking will be the subject of a future paper. There we will address the issue of heavy mode decoupling as it relates to the KK-extension of the Coleman-Weinberg model and the applicability of Symanzik's theorem.

\bibliography{KK-Decoupling-Article-Bib}

\end{document}